\documentclass[a4paper]{jpconf}
\usepackage{graphicx}
\usepackage{iopams}

\begin{document}

\title{Revisiting the Fermi Surface in Density Functional
Theory}
 
\author{Mukunda P Das$^1$ and Frederick Green$^2$}

\address{$^1$ Department of Theoretical Physics, RSPE,
The Australian National University, Canberra, ACT 2601, Australia}

\address{$^2$ School of Physics, University of New South Wales,
Sydney, NSW 2052, Australia}

\ead{mukunda.das@anu.edu.au}

\begin{abstract}
The Fermi surface is an abstract object in the reciprocal space of a crystal
lattice, enclosing the set of all those electronic band states that are filled
according to the Pauli principle. Its topology is dictated by the underlying
lattice structure and its volume is the carrier density in the
material. The Fermi surface is central to predictions of thermal, electrical,
magnetic, optical and superconducting properties in metallic systems. Density
functional theory is a first-principles method used to estimate the
occupied-band energies and, in particular, the iso-energetic Fermi surface.
In this review we survey several key facts about Fermi surfaces in complex
systems, where a proper theoretical understanding is still lacking.
We address some critical difficulties.
\end{abstract}

\section{Introduction}

Density Functional Theory (DFT) is the ``model of choice'' for
understanding condensed matter at low energies. It has achieved a
certain status as a standard first-principles method.
Indeed for many, though not all, significant condensed-matter phenomena
DFT is a powerful analytic tool
for studying electronic, vibrational, magnetic, superconducting among others.

The basis of DFT (for example, see references \cite{1}-\cite{6})
rests upon two foundational theorems by Hohenberg and Kohn \cite{7}.
Over the last five decades great progress has been made on both the
theory's fundamental aspects and its scope in application to various
systems. It has a reputation as one of the most successful practical
methods for treating many-body systems on a fully
quantum-mechanical footing.

In physical terms, however, it is many-body microscopics that ultimately
fixes the complex solid-state properties one seeks to explain and predict.
This essential many-body aspect is not addressed by DFT in any frontal
way. Its approach to many-body interactions is an implicit,
and thereby indirect, one by its positing certain well-defined but
nevertheless approximate ``exchange-correlation'' (XC) forces
between electron pairs.

The classic Kohn--Sham version of DFT \cite{8} generates effectively
single-particle simulations of what are, in actuality, interacting
correlated multi-particle systems. The theory's distinctive property
is to recover, in principle, the real-space single-particle-number
and energy densities of such a condensed system. 
Anything more than that cannot be inferred from the theorem, notably
any phenomena involving excited states beyond the single-particle
configuration. That is because such effects are
irreducibly many-body in nature and, logically, demand an explicit
many-body analysis.

To reinforce this fundamental point we recall the key physical reason
why many-body mechanisms must, at some stage, appear explicitly in any
description of interacting carriers. Landau, in his account of the weakly
interacting Fermi liquid (see for instance Nozi\`eres and Pines \cite{NP}),
showed that any energy change of an excited system away from
the ground state takes the inevitable form

\begin{equation}
\delta E[f + \delta f] \equiv
\sum_{\bf k} \varepsilon_{\bf k}[f] \delta f_{\bf k}
+ {1\over 2}\sum_{{\bf k},{\bf k'}} F_{{\bf k},{\bf k'}}[f]
\delta f_{\bf k} \delta f_{\bf k'}
\label{dft0}
\end{equation}

\noindent
where $f_{\rm k}$ is the single-particle-state distribution function, and
$\varepsilon_{\bf k}[f]$ is the corresponding single-particle spectrum
in the interacting system. (The more general Landau formula describes
the interacting free energy at finite temperature $T$; in this case $\delta f$
may as readily represent the effect of a thermal energy change, that is
away from $T=0$, as that of any other perturbation; a situation on which we
have much more to say in the following.)
Crucially, Landau's two-particle interaction
potential $F_{{\bf k},{\bf k'}}[f]$ appears in
equation (\ref{dft0}). Such an interaction is
not representable by {\em any} single-particle functional.

Although DFT, in principle, does guarantee replication of
the exact total ground-state energy $E[f]$
over the occupied single-particle states, it cannot guarantee
\smallskip

(a) that its effective single-particle spectrum, say
${\{{\widetilde \varepsilon}_{\rm k}[n]\}}_{\bf k}$, as an
auxiliary functional of the real-space density $n({\bf r})$,
is literally the replica of the actual one
${\{\varepsilon_{\rm k}[f]\}}_{\bf k}$, and
 
(b) that it will furnish a systematic estimate for the interaction potential.
\smallskip

\noindent
Both (a) and (b) are essential to a description of the full behaviour
of the particle assembly.
For example, the physical response of the real system to any
perturbation in the single-particle distribution is
characterised directly by

\[
\chi_{{\bf k},{\bf k'}}[f] \equiv
{\delta^2 E\over \delta f_{\bf k} \delta f_{\bf k'}}
\equiv F_{{\bf k},{\bf k'}}[f].
\]

\noindent
Consequently the so-called Landau interaction parameters $F$ mediate,
directly, all essential properties such as the dielectric response,
spin susceptibility, compressibility, specific heat, and so on
\cite{NP}.

As we have remarked, the interaction parameters are
inherently two-body properties that
cannot be simulated merely in single-particle terms. 
This means that density functional theory for the ground state alone,
viewed as the definitive and optimal {\em strictly single-particle}
description of a many-body system,
is insufficient to capture many significant response properties
of a real, correlated system. Yet it is the response of a system
over the entire arsenal of experimental probes that provides the
structural information one needs to discover.

The inbuilt limitations of the Kohn--Sham formalism are inherited
by any and all DFT-based machinery relying upon it exclusively. Owing to
the essentially noninteracting, single-particle nature of the
resultant models (and notwithstanding the popularity of DFT as
a practical tool applied to actual many-body problems),
a basic question of physics remains
to be answered: How can one bring to experimental test any differences (at
the very least, in the ground-state properties) between DFT as an effective
one-body theory, and the established canon of many-body analysis?

In testing the limits of DFT the detailed measurable
properties of the Fermi surface are particularly relevant.
At the best of times, these are delicate qualities to predict.
A widespread assumption still goes largely unquestioned: namely,
that the Fermi surface, with all of its intricate topology,
is uniquely a ground-state property and, uniquely, determinable via DFT.
This cannot be the case in general, least of all
for the complex interacting structures of current interest.

For ideal free carriers in an ideally uniform
three-dimensional sample, the Fermi sphere is bounded
by an equi-energetic surface
of constant curvature, of radius $k_{\rm F}$, the Fermi momentum.
Luttinger and Ward \cite{X} showed that, in such a uniform metallic
system, the Fermi surface will survive in the
presence of inter-particle interactions.
All of the low-lying excitations of the
system live near the Fermi surface and account
for much of the system's low-temperature behaviour.

Nevertheless, the real Fermi surfaces of real materials are not
simply spheres or even quasi-spheres of at least positive
curvature everywhere. Indeed they are topologically complicated,
often multiply connected, and highly subject to the
underlying crystalline geometry and interactions. They can even come
as disconnected Fermi-surface pieces, as well as exhibiting
Fermi-surface ``nesting''. The latter can induce unusual instabilities in
certain metals: for example, at sufficiently low temperature a normal metal
may move into a charge- or spin-density-wave state or other topological phase
transition.

Present condensed-state research must deal
extensively with novel and exotic materials. Invariably these are moderately
to highly correlated in terms of their strong collective electron
interactions. It follows that the relevant physics will necessarily
display its many-body nature \cite{9}-\cite{11}. Equally it follows,
in this context, that attempts to use any single-body description
must be weighed with care.

In the next Section we outline the basic features of density functional
theory.  In view of very large growth in literature we present a brief
picture of it, keeping only the salient points for our purpose. We discuss
which ground-state properties within DFT can be trusted to give a reliable
physical estimate, as well as what it is not designed to address.
In Sec. 3 we survey various popular approximations to model the
exchange-correlation functional, and present their relative merits or
demerits. In this Section we provide a list of the most often-used computer
codes that DFT practitioners rely upon.

In Sec. 4 we revisit a basic open question, still unresolved today: Is the
Fermi surface uniquely a one-body ground-state property? From a many-body
perspective this question has to be addressed carefully at zero
temperature. Experimental observation of the Fermi surfaces from
angle-resolved photo-emission spectroscopy (ARPES), Kohn anomaly,
Shubnikov-de Haas and angle resolved magneto-resistance (ARMR) methods are
discussed briefly and contrasted with calculated results of DFT,
quasi-particle (GW) theory and DMFT.
In Sec 5 finite-temperature aspects are discussed, since all actual
experiments are done at $T > 0$.  Our Summary with conclusions is
presented in the final Section.

\section{Outline of DFT}

\subsection{Hohenberg-Kohn theorem}

The Theorem for the Energy Functional of Hohenberg and Kohn
\cite{7}
(an existence proof) states the
following. Admitting certain general assumptions, any change in the
ground-state density $n({\bf r})$ of
an interacting electron system subject to an
adiabatically changing external potential $v_{\rm ext}({\bf r})$,
remains in one-to-one correspondence with that potential.

Since both $n({\bf r})$ and the total number of particles, N,
are uniquely tied to $v_{\rm ext}$  we can readily construct the full
system Hamiltonian. From it,  $n({\bf r})$
can be calculated by solving a Schr\"odinger-like equation derived from a
variational principle. We skip the many technical details (see
references \cite{1}-\cite{6})
and proceed the popular Kohn-Sham method \cite{8}.

The total energy is given as a functional of the spatial
density distribution $n({\bf r})$ by

\begin{equation}
E[n] = T[n] + U[n] + V_{\rm ext}[n] = T[n] + U[n] +
\int \!\!n({\bf r})v_{\rm ext}({\bf r}) d{\bf r},
\label{dft1}
\end{equation}

\noindent
in which the first two terms on the right-hand side are
the total kinetic energy $T$
and total inter-particle interaction energy $U$.
The total potential energy $V_{\rm ext}$ includes the 
electron-ion electrostatic interaction, here considered to be
external to the system of mobile electrons. 

In the Kohn-Sham (KS) method we consider the kinetic energy to be
exactly equivalent to that of a system of independent (noninteracting)
electrons, with a basis of single-particle orbitals
${\{ \phi_i[n] \}}^N_{i=1}$ that are themselves functionals of the exact
density distribution for the $N$ electrons in the ground state
of the system.
The theorem guarantees these orbitals to be well-defined.
They are determined systematically, in a self-consistent fashion.

The fundamental expression from the DFT analysis 
of the total ground-state energy now becomes more explicit
in detail. In this setting we have

\begin{equation}
E[n]
\equiv
T[\phi_i] + U_{\rm H}[\phi_i] + E_{\rm xc}[\phi_i] + V_{\rm ext}[\phi_i];
\label{dft2}
\end{equation}

\noindent
the interaction energy $U$ is first resolved into
the mean-field Hartree component $U_{\rm H}$ and the exchange-correlation
component $E_{\rm xc}$ with
\[
U_{\rm H}[\phi_i] = {1\over 2} \int d{\bf r} ~v_{\rm H}({\bf r}) n({\bf r}).
\]

\noindent
The set of orbitals is obtained through the Kohn-Sham equations
(here indexed by the Brillouin-zone wavevector ${\bf k}$):

\begin{eqnarray}
{\left[ {\hbar^2 k^2\over 2m}
+ v_{\rm eff}({\bf r}) \right]} \phi_{\bf k}({\bf r})
&=& {\widetilde \varepsilon}_{\bf k} \phi_{\bf k}({\bf r}),
~~~{\rm where}
\cr
\cr
v_{\rm eff}({\bf r})
&\equiv&
v_{\rm ext}({\bf r}) + v_{\rm H}({\bf r}) + v_{\rm xc}({\bf r}), ~~~{\rm with}
\cr
\cr
v_{\rm H}({\bf r})
&=&
\int \! {e^2\over |{\bf r}-{\bf r'}|} d{\bf r'} n({\bf r'}), ~~{\rm and}~~
v_{\rm xc}({\bf r})
\equiv
{\delta E_{\rm xc}\over \delta n({\bf r})}
\label{dft3}
\end{eqnarray}

Two final steps are required to close this system self-consistently.
First we define the constitutive relation for the
density in terms of the single-particle KS orbitals:

\begin{equation}
n({\bf r})
\equiv \sum_{\bf k} \theta({\bf k}) |\phi_{\bf k}[n({\bf r})]|^2
\label{dft3.1}
\end{equation}

\noindent
writing $\theta(k) = \theta({\widetilde \varepsilon}_{\rm F}
- {\widetilde \varepsilon}_{\bf k}) $
for the carrier occupation number, equal to unity within the
occupied Fermi sphere and zero for the empty momentum
states ${\bf k}$ above it;
here ${\widetilde \varepsilon}_{\rm F}$ is the corresponding DFT estimate
of the Fermi energy.
%
Second, an explicit form has to be decided for the
exchange-correlation functional $E_{\rm xc}[\phi_i]$
so that $v_{\rm xc}$ can be constructed and fed into the KS equations.
We defer this crucial point of principle to section 3.1 below.
   
By varying the total energy with respect to the undetermined
orbitals $\phi_{\bf k}$ we obtain the
$N$ Euler-Lagrange equations for the system. They are coupled nonlinear
Schr\"odinger-orbital equations, conceptually
very similar to Hartree-Fock but containing much more
correlational input, over and above exchange.

The volume of $k$-space, enclosed by the Fermi iso-surface defined as
${\{ {\bf k}; {\widetilde \varepsilon}_{\bf k}
= {\widetilde \varepsilon}_{\rm F} \}}$,
is exactly the number $N$ of mobile electrons
physically occupying the lowest-lying states within the sample.
Conservation of carriers requires this to be an invariant property
of the closed system. In particular, any imposed external field
may distort the surface and even change its topology substantially,
but cannot alter the enclosed volume
except possibly in a phase transition that totally
re-orders the ground state.

\subsection{Meaning of the Kohn-Sham eigenenergies}

Just as with Hartree-Fock theory, the early precursor of DFT,
the total energy $E[n]$
is not the sum of all the single-carrier orbital
energies ${\widetilde \varepsilon}_{\bf k}$.
In fact, we know from the derivation that the
${\widetilde \varepsilon}_{\bf k}$
enter purely as Lagrange multipliers;
variational parameters
that are strictly artifacts from the physical standpoint.
The set of ${\widetilde \varepsilon}_{\bf k}$
is simply the set of formal eigenvalues for
the auxiliary one-body equations of DFT, whose
eigenfunctions are guaranteed only to yield the correct local density
$n({\bf r})$; nothing else.

It is the net density profile, not its auxiliary
contributions, that carries the genuine physical content of
the KS equations. While the auxiliary KS eigenvalues may generally
bear some qualitative resemblance to the true energy spectrum,
there exists no guarantee that they form a trustworthy
representation of the true single-particle spectrum.

There is one important exception to the caveat above.
The value for the DFT Fermi energy
${\widetilde \varepsilon}_{\rm F}$ is, at least in principle,
the actual Fermi level in the ground state. But this does not
mean that the microscopic {\em topology} of the Fermi surface
is at all reproduced by DFT.

We stress that, although there may be practical and even heuristic
reasons to suggest that the set
${\{ {\widetilde \varepsilon}_{\bf k}\}}_{\bf k}$
describes the ``true'' band structure, to date a basic microscopic
justification of this hypothesis remains anything but settled. See
for example references \cite{12}-\cite{14}.
 
\section{Implementation of the Kohn-Sham formulation}

The only quantity that remains to be fixed is $E_{\rm xc}[n]$,
the exchange--correlation energy functional. It is formally defined
by the adiabatic connection formula
\cite{2}, \cite{5}.
For computation, the expression for $E_{\rm xc}[n]$ has to be carefully
constructed. This is where great efforts have gone into constructing several
approximate expressions, namely the so-called LDA, GEA, GGA, hybrid
functionals, ODF, and so on \cite{4}-\cite{6}. 
We now outline their main characteristics.

\subsection{Approximations for the density functional}

(a) Local-Density Approximation (LDA): the general strategy of
local-density approximations, as also those for local-spin-density (LSD)
is to take known results for the XC potential $v_{\rm xc}[\nu]$
of a uniform system at density $\nu$
and apply them locally to an inhomogeneous
system. This, in this model, $E_{\rm xc}[n({\bf r})]$
becomes a sum of {\rm locally} homogeneous (possibly spin-dependent)
exchange-correlation energies of electrons over a small cell in real space,
with a homogeneous density $\nu$ matching the local value $n({\bf r})$. 

Given this Ansatz, the total XC energy is approximated as

\[
E^{\rm LDA}_{\rm xc}[n({\bf r})]
= \int d{\bf r} ~n({\bf r}) v_{\rm xc}[n({\bf r})];
~~ v_{\rm xc}[\nu] \equiv {\delta E_{\rm xc}[\nu]\over \delta \nu}.
\]

This formula works well when density gradients are small over
the typical range of $k_{\rm F}(r)^{-1}$, the Fermi wavelength.
Forms for $E_{\rm xc}[n]$ in LDA are often taken from
parametrisation of highly precise Quantum Monte Carlo
(QMC) calculations for the electron liquid.

(b) Gradient-Expansion Approximation (GEA):
If the density variation is not small, one can try to include
systematically the gradient corrections to the LDA expressions,
going as $|n({\bf r})|$, $|n({\bf r})|^2$, etc..
In practice, low-order gradient corrections almost never
improve the LDA results and higher-order corrections are exceedingly
difficult to calculate. In any case, for real systems the results of GEA are
worse than those of LDA
\cite{2}.

(c) Generalised Gradient Approximation (GGA):
Instead of finite-order, power-series-like gradient expansions one
can use more general functionals of $n({\bf r})$ and 
${\bf \nabla} n({\bf r})$,
which need not proceed order by order.
Such functionals assume the general form

\[
E^{\rm GGA}_{\rm xc}[n]
= \int d{\bf r}~ f( n({\bf r}); {\bf \nabla}n({\bf r}) ) 
n({\bf r}),
\]

\noindent
where the $f(n({\bf r}), {\bf \nabla} n({\bf r}))$, now
non-local, is carefully constructed by satisfying at least the leading
conservation sum-rules such as perfect-screening (of each carrier by its
exchange-correlation hole). Various fitted forms are available in the current
literature \cite{6}.

(d) Hybrid functionals: These form a set of approximate forms for the
exchange-correlation energies, incorporating a portion of the exact exchange
term via Kohn-Sham wave functions together with correlation estimates from
empirical sources.

There are many parametrised hybrid forms,
some of which are of use in atomic and
molecular calculations. One of the forms is given here:

\[
E^{\rm hybrid}_{\rm xc} \equiv
a E^{\rm exact}_{\rm x} +(1-a) E^{\rm GGA}_{\rm x}
+ E_{\rm c}^{\rm GGA},
\]

\noindent
where $a$ is an adjustable mixing coefficient.

(e) Orbital-dependent functionals: This is known as the ``third generation''
of DFT. For details, see Engels' paper in chapter 2 of reference \cite{2}.
Here, instead of just density-dependent functionals, one uses
orbital-dependent functionals. Since the orbitals will obviously
embody more microscopic information, there are several advantages
of this approach to highly correlated systems.

(f) Calculations by a sort of garden variety of techniques:
these apply both to the methodology and to functional approximations
too numerous to detail here:
the so-called VASP, SIESTA, CRYSTAL, PAW,
CASTEP, Quantum Espresso, FPLO, ABNIT. And so on.
Some of these functionals have been devised for building into
computer codes developed over many years by many people.
These codes opened many new gates to the detailed computation of
many physical quantities and were popularly adopted by large numbers of
practitioners, even when the quantities calculated were
being pushed somewhat beyond the advertised
``fitness-for-use'' of the codes.

\subsection{Successes and failures}

It is difficult to describe in simple terms not only the noted successes of
DFT but also, more to the point, the sometimes minimised cases of its
failures. There are many reviews and texts to highlight fulsomely its
manifold impressive successes.

Regarding the alleged outstanding success of the GGA,
Perdew and Kurth \cite{2}
have written that
``LSD has been so successful in SSP [solid state physics] and a small
residue of GGA nonlocality in solids does {\em not}
provide a universally better description than LSD.''

Kokko and Das \cite{15} have shown LDA and GGA do not always afford
regular systematics, making it difficult to say objectively
which one is better for ground-state properties of inhomogeneous
systems, such as 3d and 4d transition metals. Many
discussions have appeared in the past few years
regarding the successes and failures of these DFT approximations
\cite{4}, \cite{6}, \cite{16}.
The general belief seems to be that
the approximations are systematically developed,
despite their remaining always somewhat uncontrolled
and therefore placing their reliability in question. 

\section{Fermi surfaces of metals}

In keeping with the central issue of this paper, at this stage
we are able to discuss questions on the nature of
Fermi surfaces and the adequacy of their description. 

\subsection{Is the Fermi surface of a metal a ground-state property?}

It is widely held that the set of iso-energetic electron bands crossing
the Fermi level $\varepsilon_{\rm F}$ define the Fermi surface (FS); a key
quantity in understanding the electronic structure of any metallic material.
This simple intersection may fix the true locus of the FS, or it may not
(depending on how faithful is the model).
But it certainly tells nothing of the
physical properties, qualitative and quantitative
over the entire surface itself. 

The conventional way of mapping a Fermi surface is to
measure the energy-distribution curves (EDC), for distinct k-points
of the Brillouin zone, via angle-resolved photoemission spectroscopy
(ARPES) and thus to ascertain the k-locations where the bands
transect the Fermi energy.
The connection between band structure and ARPES results is
authoritatively discussed in
references \cite{17} (an especially
clear review of ARPES techniques) and \cite{18}.  

The main issue is that a metal hosting
many-electron correlations will retain a well-defined Fermi surface. In
relatively weakly correlated metals, the single-electron
(i.e. independent-particle) band structure reproduces the measured FS
reasonably faithfully. As mentioned before,
certainly for strongly correlated systems
and even when strongly correlated, the one-electron
band structure is liable to become inadequate in providing
a good picture of the FS.
From equation (\ref{dft1}) the introduction we recall
the need to compute at least the two-body
Landau interaction $F_{{\bf k}, {\bf k'}}$ at
the Fermi surface to determine the response properties.

It is then clearly necessary to check the results of the single-particle
DFT for both band structure and the energy dispersion of the quasi-particle
(that is, many-body-dominated) energies. In the Fermi-liquid picture,
absent any inter-particle interactions, the band distribution of
the carriers at $T=0$ is the Fermi-Dirac step
$f_{\bf k} = \theta(\varepsilon_{\rm F} - \varepsilon_{\bf k})$.
This noninteracting distribution is precisely the formal situation
required within KS theory.

By contrast, in the presence of interactions the would-be
step function, even at zero temperature, becomes smeared around
the Fermi energy. This is because minimisation of the total
interacting ground-state energy favours a physical
configuration in which a portion of the (otherwise noninteracting)
one-body excitations is relocated in $k$-space
from somewhat below to somewhat above the Fermi level.
In other words: an increase of kinetic energy for the
distribution, by promotion of lower-lying electrons to
slightly higher states, is more than offset by a lowering of
the collective correlation energy.

Owing to this interaction-induced rearrangement, the
originally noninteracting single-particle modes
about the FS map one-to-one to stable (but now correlated)
quasi-particle (QP) state
counterparts, whose distribution retains the step-function form.
But within the noninteracting-basis picture,
this induces the appearance of a renormalisation constant $Z_k$
at the Fermi surface; $Z_k$ is always less than unity and
rescales the jump in occupancy of the underlying {\em non-interacting}
states at the Fermi level. Its size is fixed by
the self-energy correction at the FS,
arising from the many-body correlation effects.

The self-energy is immediately related to the Landau two-particle
interaction \cite{NP}
and therefore cannot be extracted from a purely one-body analysis.
For, while the knowledge of the particle-particle interaction is
sufficient to determine the self-energy and thus its contribution
to the (true) band spectrum $\varepsilon_{\bf k}$, any model spectrum
${\widetilde \varepsilon}_{\bf k}$ is not in itself sufficient
to determine the interaction. That is the core difficulty in
securing the adequacy of DFT to describe the Fermi-surface
reconstitution, or ``renormalisation''.

We arrive at a crucial point. In its many-body setting, when described
relative to the originally noninteracting single-particle basis,
the Fermi surface is always ``renormalised'' by contributions
that are explicitly many-body in nature. The FS is deformed
away from its noninteracting profile depending on
the strength of interaction. This highlights the central fact that the
physical ground state need not at all conform to the intuitive picture
provided by effective one-body formulations such as DFT
or by its conceptual ancestor, Hartree-Fock.

For simple bulk metals the Fermi temperature
$T_{\rm F} = \varepsilon_{\rm F}/k_{\rm B}$ can be large,
on the order of $10^4$K and more.
There, it is a good approximation to treat
the distribution $f_{\bf k}$ in its zero-$T$ limit;
namely, a step function with cut-off at $\varepsilon_{\rm F}$.
However, for complex metals possessing various phase transitions
at low temperatures, the $T=0$ limiting approximation will be of
no use in capturing the correct physics. That is because the interactions
themselves, quite apart from any thermal effects,
have already smeared it out.

Even when phase transitions are not in play, we have seen how
many-body renormalisation can induce a radical change in 
the landscape of the critical states
around a ``normal'' Fermi surface. In the potential
presence of phase changes, one has even less option
than to approach head-on the theory of finite-temperature effects
on electronic structure. The quick fix does not exist here. 

Energy bands as estimated by DFT differ substantially
from the actual QP states described by many-body
theory. This is evident when comparing the respective DFT and QP
expressions

\begin{eqnarray}
{\left[ {\hbar^2\over 2m} k^2
+ v_{\rm ext}({\bf r}) + v_{\rm H}({\bf r}) + v_{\rm xc}({\bf r}) \right]}
\phi_{\bf k}({\bf r})
&=&
{\widetilde \varepsilon}_{\bf k} \phi_{\bf k}({\bf r});
\cr
\cr
{\left[ {\hbar^2\over 2m} k^2
+ v_{\rm ext}({\bf r}) + v_{\rm H}({\bf r}) \right]}
\psi_{\bf k}({\bf r})
+ \int \!d{\bf r'} \Sigma({\bf r}, {\bf r'}; \varepsilon_{\bf k})
\psi_{\bf k}({\bf r'})
&=&
\varepsilon_{\bf k} \psi_{\bf k}({\bf r}).
\label{dft4}
\end{eqnarray}

\noindent
In the latter expression arising from standard many-body analysis,
$\Sigma({\bf r}, {\bf r'}; \omega)$
is the nonlocal, irreducible electronic self-energy.
This dynamical self-energy term is far more structured
and internally complex than its static DFT analogue $v_{\rm xc}$.  

There exists a wide range of systematic and consistent approximations
for $\Sigma$. A relatively simple and successful such model is
the ``GW'' approximation, in which the ``G'' stands for the
one-body Green function and the ``W'' for the microscopically
screened two-body interaction \cite{19}.
In the frequency domain
$\Sigma({\bf r}, {\bf r'}; \omega)$
is given by

\[
\Sigma({\bf r}, {\bf r'}; \omega)
\equiv i \int d\omega' G({\bf r}, {\bf r'}; \omega + \omega')
W({\bf r'}, {\bf r}; \omega'), 
\]

\noindent
in which the screened electron-electron interaction is approximated by

\[
W({\bf r'}, {\bf r}; \omega')
\equiv  \int d{\bf r''} \epsilon^{-1}({\bf r'},{\bf r''}; \omega)
v({\bf r''} - {\bf r}). 
\]

\noindent
and here $\epsilon^{-1}({\bf r'},{\bf r''}; \omega)$ 
is the inverse of the dynamic dielectric function. 

If $v_{\rm xc}$ were a good approximation to
$\Sigma$ the DFT band energies ${\widetilde \varepsilon}_{\bf k}$
would also be good. However, comparison of
calculations performed within KS and GW show
large differences near the Fermi wave vector.
There are many examples of DFT Fermi surfaces which
are manifestly at variance with the physical, many-body QP 
Fermi surface. This is particularly so in multi-band
and correlated materials \cite{20}, \cite{21}.

\subsection{Fermi surfaces from dynamical mean-field theory}

Dynamical mean-field theory (DMFT) is a new development
that takes into account local correlations more faithfully
\cite{22}.
It is an improved GW method, where local correlations are
incorporated by treating them as effective impurities in
a periodic system. In brief, the method produces an effective mass and
the associated renormalisation for the QPs can be a substantial,
consistent with observed ARPES \cite{23} in correlated electron systems.

\subsection{Is the DFT-Kohn-Sham Fermi surface a ground-state property?}

At the start of this Section we examined the wider issues
of principle leading to this question. The answer
is straightforward. The KS equations
certainly reproduce the correct physical density distribution and
the total ground-state energy. In the course of things, given
that a band structure emerges out of KS, a FS must also
exists for it.

Nevertheless, when it comes to real systems with their
interactions and inhomogeneities,
the FS of DFT-KS is already known to be inequivalent
{\em in principle} to the physical FS obtained
from the microscopic Dyson equation \cite{24}, \cite{25},
which codifies the many-body effects within the true QP distribution.
A detailed theoretical argument demonstrates that the
inaccuracy of the FS in DFT-KS theory comes from an
inbuilt lack of sufficiently strict convergence for the
gradient approximation implicit in DFT.

In an analysis using a time-dependent generalisation of DFT,
Cohen and Wasserman
\cite{CW}
conclude that KS FS is identical to the QP FS in the sense
of a ``distributional'' argument -- that is, at best in some
average sense.

\subsection{Are explicit many-body effects seen in the FS properties
of metals?}

We cite some apposite and fruitful observations on the issue.

\begin{itemize}
\item A.K. Rajagopal notes in his review \cite{12}:
``The eigenvalues [of the DFT equation] do not have any special
significance. The equation is a mathematical artifact of the HKS
[Hohenberg-Kohn-Sham] formalism. By taking a pragmatic point of
view by treating $E_i$ [our ${\widetilde \varepsilon}_{\bf k}$]
as a one electron eigenvalue in the one electron
theory of band structure, one arrives at the HKS band structure.
There is much controversy regarding the definition
of ``Fermi surface'' whether it is a ground state property
of the system or not?''

\item Richard Martin, on p. 131 of his text \cite{13},
also visits this point. We quote:
``Is the exact Fermi surface of a metal given by the exact ground
state DFT? .... this is not a trivial question for two reasons (i) a
many-body metal must have a well-defined FS - this is assumed for the
purpose.  (ii) It is not a-priori obvious that FS is a ground state
property. One way to see if the FS is determined by a ground state property
is to consider the susceptibility to static perturbations. The exact DFT must
lead to the correct Kohn anomaly and Friedel oscillations of the density far
from an impurity, which depend in detail on the shape of the FS of the
unperturbed metal.''
\end{itemize}

We return to Martin's last point in detail, in subsection 4.7
below; since a singularity in the dielectric
susceptibility $\chi(q)$ is induced by electron-hole FS excitations
when $q = 2k_{\rm F}$ spans the Fermi surface (the Kohn anomaly),
one may reconstruct the physical FS from this information.
The loci of the singularity
trace out the FS. It is thus natural, and appropriate, to
interrogate the effectiveness of DFT-based approaches
as to how well their FS estimates match measured singularity data.

\subsection{Instability of the Fermi surface}

If the FS is a robust ground-state property, exactly
how stable is it against perturbations? Here are several examples.

\begin{itemize}
\item For a number of reasons including interactions, temperature
and impurities, a system of metallic electrons can undergo a phase
transition precipitated by superconducting or magnetic pairing;
by a charge- or spin-density wave (CDW/SDW); by a
nematic state; or a number of other many-body collective states.

\item
As mentioned in the Introduction, Landau's theory of the Fermi liquid
of 1957 became the canonical description of the metallic state. It was
realised, even at that time, that the FS is unstable against strong
interactions; the Pomeranchuk instability. A simple example occurs when
an initially isotropic FS loses its symmetry in the presence of a strong
interaction. The system will minimise its free energy at the cost of
the FS becoming distorted and anisotropic. Again, the gain in kinetic
energy is more than offset by a lowering of correlation energy.

\item
During the past few years a number of models in 2D systems have
been studied, both theoretically and experimentally. These have
once more brought to the fore important questions in our understanding
of the FS as an invariant ground-state property
\cite{26}-\cite{31}.
\end{itemize}

\subsection{How does one determine the FS?}

A good exposition of Fermi-surface analysis is
given in the text by Ziman \cite{31}.
Currently one can try to map out the FS by the following means:

(i) Theoretically, by DFT through the naive KS band structure
(with formally questionable status).

(ii) DFT augmented with {\em exact} response theory. 

(iii) Empirically, by ARPES \cite{17}, \cite{18} measurements:
the ARPES spectrum is not just a set of one-electron bands,
but measures directly quasi-particle spectral function $A({\bf k},\omega)$,
given theoretically by the imaginary part of the Green function
for one-particle excitations, with interaction effects
(self-energy) fully included:

\[
A({\bf k}, \omega) = -{1\over \pi}{\rm Im}
{\left\{ G({\bf k}, \omega) \right\}}
= -{1\over \pi} {\rm Im} {\left\{ { 1\over
{\hbar \omega - \varepsilon^0_{\bf k} - \Sigma({\bf k},\omega)} }
\right\}}
\]

\noindent
where $\varepsilon^0_{\bf k}$ is the noninteracting single-carrier
energy and $\Sigma({\bf k},\omega)$ can be calculated by GW or DMFT.

(iv) Magnetic resonance effects. Electronic Fermi surfaces are selectively
measured by observing of the oscillation of transport properties in
differently oriented magnetic fields ${\bf H}$. This approach
results in, for example, the de Haas--van Alphen effect (dHvA) and the
Shubnikov--de Haas effect (SdH). The former is an oscillation in magnetic
susceptibility and the latter in resistivity. The determination of the
periods of oscillation for various strengths and directions
of ${\bf H}$ allows one to infer the size and shape of the
Fermi surface \cite{31}.

(v) Mapping out the FS using Angle-Dependent Magneto-Resistance
oscillations (ADMRO) \cite{33}, \cite{34}. Reference \cite{33}
has a comparison of ARPES and the ADMRO measurements in
${\rm Sr}_2{\rm RuO}_4$. The authors point out certain
obvious inconsistencies in the ARPES data.

\subsection{Fermi surface from canonical response theory}

The Lindhard formula for the charge-density response
(see reference \cite{31}, p 129)
is defined for noninteracting electron states,
with occupancy $f^0_{\bf k}$
at band energy $\varepsilon^0_{\bf k}$.
The dielectric function $\epsilon({\bf q}, \omega)$
of an electron liquid is then given by

\begin{equation}
\epsilon({\bf q}, \omega) = 1 - v(q) \chi({\bf q}, \omega)
~~{\rm where} ~~
\chi({\bf q}, \omega) = \sum_{\bf k}
{ {f^0_{{\bf k} + {\bf q}} - f^0_{\bf k}}\over
{\hbar \omega - \varepsilon^0_{{\bf k} + {\bf q}}
+ \varepsilon^0_{\bf k}} }.
\label{dft5}
\end{equation}

\noindent
Evaluating this quantity and taking its static limit
$\omega \to 0$ at finite $q$ and $T=0$,
we have

\begin{equation}
\chi({\bf q}) \equiv \chi({\bf q}, 0)
=
{3n\over 4\varepsilon_{\rm F}} {\left[
1 + {{4k^2_{\rm F} - q^2}\over 4k_{\rm F} q}
\ln{\left| {{2k_{\rm F} + q}\over {2k_{\rm F} - q}} \right|}
\right]}.
\label{dft6}
\end{equation}

\noindent
This formula is logarithmically singular at $q=2k_{\rm F}$. 

The dielectric function is continuous but its $q$-derivative has a
logarithmic infinity at $q=2k_{\rm F}$. A more realistic calculation
shows that the singular logarithmic behaviour is not only evident
in the static dielectric susceptibility for an isotropic FS,
but indeed that the logarithmic anomaly occurs for any Fermi surface, where
$\chi({\bf q}, \omega)$ assumes the quite general form displayed
in the second of the expressions given in equation (\ref{dft5}).

The Kohn singularity will manifest itself under very broad conditions.
It demands only that there be a non-vanishing
iso-energetic Fermi surface.
If one selects a measurement with ${\bf q} = 2{\bf k}$
as a function of orientation for those Fermi momenta
spanning the Fermi surface (that is, such that
$\varepsilon_{\bf k} = \varepsilon_{\rm F}$)
one maps out the physical FS in reciprocal space.

For a lattice system the inter-ionic potential is quasi-statically
screened by the dielectric function. As a result the phonon frequency,
which is much lower than the Fermi energy, is itself
dependent on $\epsilon({\bf q}) = 1 - v(q)\chi({\bf q})$.
Thus the singularity is reflected as a distinct
``kink'' in the phonon dispersion, known as the Kohn anomaly
\cite{KA}.

The dynamic dielectric function  $\epsilon({\bf q},\omega)$
can be constructed from detailed many-body theory by systematic
inclusion of the exchange-correlation forces.
In this context two theories are at hand: (1) the Kohn-Sham
theory of independent electrons (subsuming all many-body effects into
an effective one-body potential) and (2) explicit many-body terms
contributing directly to the Dyson equation for the
quasi-particle excitations.
One can then very well see that the Kohn singularity
appears in both KS and QP Fermi surfaces, since both approaches
lead to a well-defined estimate for it. They may then be compared
as to their agreement with the Kohn-anomaly phonon data, among others.

\section{Finite-temperature effects on the FS}

DFT was formally extended by Mermin \cite{34} to include finite temperatures,
by considering a grand canonical ensemble of particles.
Unlike the case of the
theory for $T=0$, no rigorous DFT functional is available for the grand
potential. Nevertheless there are some approximate LDA-type functionals that
have been invoked in plasma physics, nuclear physics and quantum-chemistry
applications of DFT
\cite{35}.

To our knowledge, finite-temperature band structures are rarely considered
relevant to metal physics. This is understandable for simple metals,
since the Fermi temperature far exceeds the scale of any experimental $T$.
The electronic structure and FS at room temperature and
below are reasonably well understood, taking into account the constraints
previously discussed.  Further, these systems do not exhibit any phase change
at low temperatures. Here, therefore, one believes that $T=0$ DFT is adequate
and there are many calculations reported in the literature.

In reference \cite{36} a calculation by the LDA+DMFT method is compared
with ARPES and dHvA experimental results. It is shown there that
DFT calculations, either with LDA or GGA, fail to reproduce the
experimentally observed electronic structure of the multi-band material
${\rm KFe}_2{\rm As}_2$.

In this report our main attention falls on metals of multi-band type, with
multiple FSs and many phase transitions at low temperatures.
Fermi-surface instabilities at finite temperature form a
topic of intense interest as of this writing
\cite{37}.
Typical examples are  ${\rm Sr}_2{\rm Ru}_2{\rm O}_7 $
and many crystalline pnictides
\cite{38}--\cite{40}.
In such cases, finite-temperature effects ought to be considered
within the relevant theories to be applied.
Since serious finite-temperature DFT has not been attempted -- in contrast
with rather naive extrapolations of DFT at $T=0$ -- many issues have now
come to attention.

We first remark that $k_{\rm F}$ ceases to be a relevant parameter
away from $T=0$, for cases in which the condition
$\varepsilon_{\rm F} \gg k_{\rm B}T$ ceases to hold.
Rather, when Fermi and thermal energies become comparable,
a different wave-vector $k_m$ comes into play,
defined in terms of the actual chemical potential $\mu(T)$ by

\[
{{\hbar^2 k_m^2}\over 2m^*[\mu(T)]} + U(k_m) \equiv \mu(T)
\]

\noindent
where $U$ is the momentum-dependent mean-field potential.
For low temperature the sharpness of the FS still exists but
only so long as $k_{\rm B}T \ll \mu$.

There is already a remarkable example of a large finite-temperature effect
that is inaccessible to simplistic extrapolations of zero-temperature
methodology. Backes et al.
\cite{41}
performed a finite-temperature and -pressure DFT based on
Born-Oppenheimer molecular dynamics for a Fe pnictide system.
Their finite-$T$ results, in pronounced contrast with the
electronic structure and FS computed at zero temperature,
are new and entirely different
\cite{41}.
The authors report that the band structure as well as the density of states,
unlike those for $T=0$, exhibit pronounced structural oscillations
when measured at finite $T$. Oscillation of the band energies,
particularly near the Fermi surface at ambient pressure, then
leads to thermal broadening as expected.

The investigations by Backes et al. show that the FS at finite
temperature has a conspicuously different behaviour in experimental
observations, substantially at odds with DFT-based estimates relying
on purely zero-temperature arguments.
These need to be investigated further. Given these experimental
findings, the finite-temperature aspects are undoubtedly missed
by any effectively $T=0$ density functional calculation.

In contrast with the Backes et al. work, a recent
instance of a conventional DFT calculation taken outside its
proper limit of validity is provided by Sen et al.
\cite{Sen},
who have advanced a misleading picture of the FS structure for
FeAs-based pnictide materials at finite temperature.
Their estimate is improperly generated in that
what should be, inherently, a fully finite-temperature
calculation is treated by an unjustified
extension away from a strictly $T=0$ basis.
We will expand this issue in a follow-up paper.

\section{Conclusions}

This brief overview deals with several nontrivial issues related to the
ground-state predictions of density functional theory for metals. In
particular we have reviewed what is needed for a clear and more faithful
characterisations of the Fermi surface.

The FS is a central property of conducting electron systems, necessary
for understanding their many physical properties. Transport and
superconductivity are but two of them among a wide set of interesting and
important many-body phenomena.

We started with a concise overview of the DFT, recognised as a
first-principles theory for many-body systems. Since no real many-body
problem can be solved exactly, the required systematic approximations
to the many-body exchange-correlation functionals are outlined.
Many computer codes, both general and specialised, rely on these
approximations and are in widespread use in calculating almost every
property, at least at zero temperature.

We have highlighted the Fermi surface as a special and especially
delicate property of a Fermi system of carriers, and have looked at
how one can use various experimental methods to map it:
ARPES, SdH, dHvA, ARMRO and the interesting Kohn-anomaly
technique. Since any measurements are carried out at finite temperature,
we have emphasised the need for proper applications of finite-temperature
theory rather then a too-casual stretching of strictly zero-temperature
results. This becomes particularly evident when the Fermi energy of a
metallic or semi-metallic material starts to enter the typical
thermal scale. Uncritical extrapolations of $T=0$ results cannot but
fall short of physical accuracy. At best (in the circumstances relevant
to contemporary investigations) the claim of 
first-principles reliability cannot be supported in their case.

The importance of finite-temperature theory is argued in the context of some
novel correlated electron systems, notably for those with phase changes at
low temperatures. We believe there are some important aspects of the physics
of ``fermiology'' that need serious extension, in not indeed a thoroughgoing
re-evaluation.

Finally, we emphasise that for a number of crucial material properties,
density functional theory has proved itself one of the most fruitful
and rigorous tools for furthering progress in condensed-matter physics
over the last half-century. Our aim here is far from criticising
the theory or belittling its manifold deep successes. Density functional
theory, as with all theoretical innovations, has limits. It is rather
with a view to upholding the true strengths of DFT that we have
visited some delicate, nonetheless themselves crucial, aspects of the
response\ of materials where the potential for uncritical misapplication of
density functional arguments poses certain novel conceptual problems.

\section*{Acknowledgment}

We thank Professor Reiner Dreizler for helpful correspondence.

\section*{References}

\end{document}